\documentclass{article}
\usepackage{epsfig}
\begin{document}
{\centerline{\Large\bf
A dual view of the 3d Heisenberg model}}
{\centerline{\Large\bf
and the abelian projection.}}
\vskip0.3in
{\centerline{
{\large A. Di Giacomo$^a$, D. Martelli$^b$ and G. Paffuti$^a$}}}
\vskip\baselineskip
\begin{itemize}
\item[a)] Dipartimento di Fisica Universit\`a and INFN Pisa (Italy)
\item[b)] ISAS and INFN Trieste (Italy)
\end{itemize}
\vskip0.5in
{\centerline{\large\bf Abstract}}
The Heisenberg model in 3d is studied from a dual point of view. It is shown
that it can have vortex configurations, carrying a conserved charge ($U(1)$
symmetry).

Vortices condense in the disordered (demagnetized) phase. A
disorder parameter $\langle\mu\rangle$ is defined, dual to the
magnetization $\langle\vec n\rangle$, which signals condensation
of vortices, i.e. spontaneous breaking of the dual $U(1)$
symmetry. This study sheds light on the procedure known as abelian
projection in non abelian gauge theories. \vskip0.5in {\tt PACS:
11.15.Ha, 64.60.Cn} \vfill\eject
\section{Introduction}
Order disorder duality\cite{1,2} plays an increasingly important role in our
understanding of gauge theories, specifically of QCD\cite{3} and of its
supersymmetric extensions\cite{4}.

Duality exists in systems which have spatial configurations with non trivial
topology, carrying a conserved topological charge. It consists in the
possibility of describing the system in terms of two sets of fields: a set of
local fields which is convenient to describe the ordered phase at weak coupling
(low temperature),
and a dual set which is convenient at strong coupling (high temperature).
The non zero vacuum expectation value (vev) of a local field
(order parameter) $\langle \vec\Phi\rangle$ signals order. At some value of the
coupling (temperature) $\langle \vec\Phi\rangle\to 0$ and there is a transition
to disorder. In the disordered phase topological excitations condense, the
symmetry corresponding to conservation of the topological charge is
spontaneously broken. In terms of the dual variables the system looks ordered,
and the non zero vev of some operator $\mu$, $\langle\mu\rangle$ carrying non
zero topological charge is the dual order parameter.
When expressed in terms of the original fields $\mu$ is a highly non local
operator.
The dual order parameter is
usually called a disorder parameter.

The prototype system showing duality is the 2d Ising model. The
field  is a two valued variable $\sigma(n) = \pm 1$, defined on
the sites of a 2d square lattice. At low temperature there is
order and $\langle\sigma\rangle \neq 0$, breaking the symmetry
$\sigma\leftrightarrow -\sigma$. A transition point exists where
$\langle\sigma\rangle \to 0$; at higher temperatures the system
looks disordered. However the same system can be described in
terms of a dual variable $\sigma^*(n^*) = \pm 1$, again on a
square lattice which roughly speaking associates a site to each
spatial link of the original lattice with the rule that
$\sigma^*=-1$ if the values of $\sigma$ at the ends of the link
are equal and $\sigma^* = 1$ if they are different.

Topologically non trivial one dimensional (spatial) exicitations exist, which
in terms of $\sigma$ look as kinks (half space with $\sigma=-1$, half space
with $\sigma=1$) and are highly non local. In terms of $\sigma^*$ a kink is a
configuration with a single spin up. When described in terms of $\sigma^*$
(dual variable) the system is again an Ising model: it is self dual.

The only change is that the Boltzman factors, $\beta$ and $\beta^*$, of the two
descriptions are related by
\begin{equation}
 \sinh\frac{2}{\beta} = \frac{1}{\sinh\frac{2}{\beta^*}}\label{eq:int1}
\end{equation}
Ordered region of one description correspond to disordered region of the other
and viceversa. In the disordered phase $\langle\sigma\rangle = 0$,
$\langle\sigma^*\rangle \neq 0$. $\langle\sigma^*\rangle$ is the disorder
parameter.

A similar situation occurs in the 3d XY model\cite{5} (liquid He$_4$) where 2d
configurations (vortices) with a conserved quantum number exist. In this case
the dual system is a Coulomb gas\cite{6} and there is no self duality.

In the 4d compact $U(1)$ gauge theory the 3d configurations with topology are
monopoles, which condense in the disordered phase producing confinement of
electric charges\cite{7}.

Not always in these systems it is possible to explicitely write the partition
function in terms of the dual variables.

An alternative approach is to write the disorder operator in terms of the
original fields\cite{8} (not their dual). This is particularly practical in
numerical simulations. A (non local) operator is written which creates a
topological excitation, and therefore carries non zero topological charge, and
its vev is measured, in order to detect spontaneous breaking of the dual
symmetry\cite{5,9}.

In the case of QCD there exists no clear idea of what the dual description is,
except that it is presumably a gauge theory, with a gauge group independent of
the colour group, and an interchange of roles between electric and magnetic
charges\cite{7,10}.

The topological excitations which condense should have magnetic charge. This
supports the idea that confinement is produced by dual Meissner effect in a
dual superconducting vacuum\cite{11}.

Monopoles exist in QCD in connection with any operator $\Phi$ in the adjoint
representation: they are exposed by a procedure known as abelian
projection\cite{12}.

A disorder parameter can be defined, wich detects condensation of the monopoles
for each operator $\Phi$. Numerical simulations do demonstrate that monopoles
condense inthe vacuum, independent of the choice for $\Phi$\cite{13,13a}.

In this paper we want to study the 3d ferromagnet Heisenberg model in a dual
way.
The motivation for that, besides its intrinsic interest, is twofold
\begin{itemize}
\item[1)] Have another example of duality and a check of the construction of
the disorder parameter.
\item[2)] Have some insight into the abelian projection.
\end{itemize}
The model presents topological configurations in 2 dimensions, the well known
instantons of the 2d $O(3)$ $\sigma$ model\cite{14a,15}.

We will numerically check that the disordered (high temperature phase) of the
model is a condensate of such excitations, by use of a non local disorder
parameter.

The idea will prove correct and as a check finite size scaling
analysis at the Curie point will provide a determination of the
critical indices and of the transition temperature, which agree
with the ones produced by other methods.

To make connection with gauge theory we shall reformulate the
model in terms of a fiber bundle. We shall also consider a gauged
version of it, which is a $2+1$ dimensional Georgi-Glashow model.
The Heisenberg model can be viewed as the limit of zero gauge
coupling of it. The topological charge turns out to be the
corresponding limit of t'Hooft's magnetic charge.

The paper is organized as follows.
In sect.2 we define the model, and formulate it in terms of a fiber bundle. In
sect.3 we define the disorder parameter, and present its numerical evaluation,
together with the determination of the transition point and of critical
indices. Some remarks on the construction of the disorder parameter are
contained in
sect.4. Sect.5 describes the gauged version and discusses the limit of zero
gauge coupling. Sect.6 contains some concluding remarks.

\section{The model.}
The partition function is
\begin{equation}
Z[\beta] = \int\prod_x[d\Omega(x)]\,\exp(-S)
\label{eq:2.1}\end{equation} with  $d\Omega(x)$ the area element
on the unit sphere in colour space and
\begin{equation}
S = \frac{1}{2}\beta\sum_{\mu=0}^2\sum_x (\Delta_\mu\vec n(x))^2
\qquad \Delta_\mu\vec n(x) =
[\vec n(x+\hat \mu) -\vec n(x)]\label{eq:1.a1}
\end{equation}

A generic $x$ dependent $O(3)$ transformation
\begin{equation}
\vec n(x) \to U(x)\,\vec n(x)\label{eq:2.2}\end{equation}
leaves $Z[\beta]$ invariant, in spite of the fact that $S$ is not invariant,
since it can be reabsorbed in a change of the variable $\Omega(x)$ which leaves
the measure $d\Omega(x)$ invariant.

The continuum version of the model is the nonlinear $O(3)$
$\sigma-$model\cite{1}
\begin{equation}
{\cal L} = \frac{1}{2}(\partial_\mu\vec \sigma)^2\qquad
(\vec\sigma^2 =1)\label{eq:2.3}\end{equation}
A gauged version of the model is in the continuum
\begin{equation}
{\cal L} = \frac{1}{2} (D_\mu \vec \sigma)^2 -\frac{1}{4}\vec
G_{\mu\nu}\cdot\vec G_{\mu\nu} \label{eq:2.4}\end{equation}
\[ D_\mu = \partial_\mu - i g \vec T\cdot\vec A_\mu\qquad
\vec G_{\mu\nu} = \partial_\mu \vec A_\nu - \partial_\nu\vec A_\mu +
g\vec A_\mu\wedge\vec A_\nu\]
which is a $2+1$ dimensional Georgi-Glashow model, with the length of the Higgs
field frozen to 1.

The Heisenberg model can be considered as the limit $g\to 0$ of
the system (\ref{eq:2.4}).

Usually an $x$ independent frame is used in colour space, or  3 fixed  unit
vectors
$\vec\xi^0_i$ ($i=1,2,3$) with
\begin{equation}
\vec\xi^0_i\cdot\vec\xi^0_j=\delta_{ij}\qquad
\vec\xi^0_i\wedge\vec\xi^0_j = \vec\xi^0_k\label{eq:2.7}\end{equation}
However a body fixed frame (BFF) can be used\cite{16}, with unit vectors
$\vec\xi_i(x)$ ($i=1,2,3$) again obeying
\begin{equation}
\vec\xi_i\cdot\vec\xi_j=\delta_{ij}\qquad
\vec\xi_i\wedge\vec\xi_j = \vec\xi_k\label{eq:2.8}\end{equation}
and $\vec\xi_3(x) = \vec n(x)$.

The frame is determined modulo an $x$ dependent arbitrary rotation around $\vec
n(x)$.

Since $\vec\xi_i^2 = 1$
\begin{equation}
\partial_\mu\vec\xi_i(x) = \vec\omega_\mu\wedge\vec \xi_i(x)
\label{eq:2.9}\end{equation}
or
\begin{equation}
D_\mu\vec \xi_i(x) = 0 \label{eq:2.10}\end{equation}
where we have defined the covariant derivative
\begin{equation}
D_\mu \equiv \partial_\mu - \vec\omega_\mu\wedge = \partial_\mu - i \vec
T\cdot\vec\omega_\mu\label{eq:2.11}\end{equation}
Eq.(\ref{eq:2.10}) also implies that
\begin{equation}
\left[ D_\mu(\omega), D_\nu(\omega)\right]\vec \xi_i(x) = 0
\label{eq:2.12}\end{equation}
or, by the completeness of $\vec \xi_i$
\begin{equation}
\left[ D_\mu(\omega), D_\nu(\omega)\right] = \vec T\cdot \vec G_{\mu\nu} = 0
\label{eq:2.13}\end{equation}
$\vec \omega_\mu$ is a pure gauge
\begin{equation}
\partial_\mu \vec \omega_\nu - \partial_\nu\vec \omega_\mu +
\vec \omega_\mu\wedge\vec \omega_\nu = 0 \label{eq:2.14}\end{equation}
The geometrical meaning of the above analysis is nothing but the very
definition of parallel transport. Eq.(\ref{eq:2.14}) is true, apart from
singularities.
A consequence of (\ref{eq:2.14}) is that any field configuration $\vec n(x)$
can be written as a parallel transport from infinity to the point $x$ of the
value $\vec n_0$ which the field has at
some point at
infinity
\begin{equation}
\vec n(x) = {\cal P}
\left(\exp i\int_{\infty,{\cal C}} \vec T\cdot\vec\omega_\mu(x) \, d
x^\mu\right)\vec n_0 \equiv R(x)\vec n_0\label{eq:2.15a}\end{equation}
${\cal P}$ means path-ordering. The choice of the path ${\cal C}$ is irrelevant
as long as the field $\vec \omega_\mu$ is a pure gauge, or eq.(\ref{eq:2.14})
is satisfied.
We will show that
this is not true, because singularities of
the matrix $R$ in eq.(\ref{eq:2.15a}), which make the connection of the bundle non
trivial.

We now analyse the nature of such singularities.

The model has a No\"ether current
\begin{equation}
\vec J^\mu_{(N)} = \vec n\wedge\partial^\mu\vec n\label{eq:2.15}\end{equation}
corresponding to $O(3)$ invariance which is conserved by virtue of the equation
of motion
\begin{equation}
\partial_\mu J^\mu_{(N)} = 0 \label{eq:2.16}\end{equation}
There is, however another current which is  identically conserved
\begin{equation}
\vec J_\mu = \frac{1}{8\pi} \varepsilon_{\mu\alpha\beta}\partial_\alpha\vec
n\wedge\partial_\beta \vec n \label{eq:2.17}\end{equation}
$\vec J_\mu$ is parallel to $\vec n$, since both $\partial_\alpha\vec n$ and
$\partial_\beta\vec n$ are orthogonal to it.
\begin{equation}
\partial^\mu \vec J_\mu = 0\label{eq:2.18}\end{equation}
Also the colour invariant current
\begin{equation}
J_\mu = \vec n\cdot\vec J_\mu =
\frac{1}{8\pi} \varepsilon_{\mu\alpha\beta}\vec n\cdot\partial_\alpha\vec
n\wedge\partial_\beta \vec n\label{eq:2.19}\end{equation}
is identically conserved
\begin{equation}
\partial^\mu J_\mu = 0
\label{eq:2.19a}\end{equation}
The conservation law (\ref{eq:2.19a}) is a consequence of the invariance with
respect the $x$ dependent rotations around $\vec n(x)$. The action (\ref{eq:2.3})
can indeed be written as
\[
S = \frac{\beta}{2}\left(\vec \omega_\mu\wedge\vec n\right)^2 =
\frac{\beta}{2}(\vec\omega_\mu^\perp)^2\] Any rotation around
$\vec n$ corresponds to an $\vec \omega$ parallel to $\vec n$, and
leaves $S$ invariant. In defining the BFF this invariance reflects
in the arbitrariness by a rotation in the choice of $\vec\xi_1$,
$\vec\xi_2$, being $\vec\xi_3$  parallel to $\vec n$.

The conserved charge corresponding to (\ref{eq:2.19a}) is
\begin{eqnarray}
Q &=& \frac{1}{4\pi}\int\vec n\cdot(\partial_1\vec n\wedge\partial_2\vec n)
d^2 x\label{eq:2.22}\\
&=& \frac{1}{4\pi}\int\vec n\cdot(\vec\omega_1\wedge\vec\omega_2)\,d^2 x
\nonumber\end{eqnarray}
$Q$ is nothing but the Chern number of the $2d$ $O(3)$ $\sigma-$model, which
assumes positive or negative integer values,
\begin{equation}
Q = \pm n\label{eq:2.23}\end{equation}
On the other hand, by use of eq.(\ref{eq:2.14})
\begin{eqnarray*}
Q
&=&\frac{1}{4\pi}\int\vec n\cdot(\vec\omega_1\wedge\vec\omega_2)\,d^2 x=
\frac{1}{4\pi}\int(\partial_1\vec\omega_2\cdot\vec n -
\partial_2 \vec\omega_1\cdot\vec n) d^2 x =\\
&=&\frac{1}{4\pi}\int(\partial_1(\vec\omega_2\cdot\vec n) -
\partial_2 (\vec\omega_1\cdot\vec n)) d^2 x +
2\frac{1}{4\pi}\int\vec n\cdot(\vec\omega_1\wedge\vec \omega_2)\,d^2x
\end{eqnarray*}
and hence, since the last term equals $2 Q$,
\begin{equation}
\pm n =
\oint\vec\omega_i\cdot\vec n dx^i\label{eq:2.24}\end{equation}
showing that the field $\vec \omega_\mu$ can have  a nontrivial
connection.
There exist singularities of $\vec \omega_\mu$ where
eq.(\ref{eq:2.14}) is not valid.

This can be seen by expressing $\vec\xi_i$ in polar coordinates with respect to
$\vec\xi_i^0$, the fixed frame axes. Then one can compute $\vec \omega_\mu$
getting\cite{16}
\begin{equation}
\vec\omega_\mu =
\left(\matrix{\sin\theta \partial_\mu \psi\cr
-\partial_\mu \theta\cr
-\cos\theta\partial_\mu \psi\cr}\right)
\label{eq:2.25}\end{equation}
Singularities can occur when $\cos\theta=\pm1$, and $\psi$ is not defined.
The singularity is then of the form
\[
\vec\omega^{sing}_\mu =
\left(\matrix{0\cr
0\cr
\pm\partial_\mu \psi^{sing}\cr}\right)
\]
The corresponding field is an abelian field parallel to $\vec n$ in the sites
where $\vec n = \vec n_0$
\[ \vec F^{sing}_{\mu\nu} =
\pm\vec n_0(\partial_\mu\partial_\nu - \partial_\nu\partial_\mu)\psi^{sing}
\]
The singular field can be explicitly computed for a time independent soliton
solution (a 2d instanton independent of $x_0$), sitting at the origin, getting
(see sect.5)
\[ \partial_1\omega_2^{sing} - \partial_2 \omega_1^{sing} =
2\pi\delta^{(2)}(\vec x)\]
This is nothing but a Dirac string.
\section{The disorder parameter}
We will show that solitons condense in the disordered phase of the model
$(\beta < \beta_c$), i.e. that the $U(1)$ symmetry eq.(\ref{eq:2.19a}) is
spontaneously broken in the disordered phase. To do that we define a disorder
parameter which is the vacuum expectation value (vev) of the creation
operator of a soliton.

We start by defining the creation operator of a soliton.

Let $R_q(\vec x,\vec y)$ be a singular time independent rotation which creates a
soliton of topological charge $q$
at site $\vec y$. We will give and discuss the explicit form  of
$R_q$ below.

We define the lattice creation operator of a soliton at site $\vec y$ and time
$t$ as follows
\begin{eqnarray}
\mu_q(\vec y,t) &=&
\exp\Bigl\{
-\beta\Bigl[
\sum_{\vec x}( R^{-1}_q(\vec x,\vec y)\vec n(\vec x,t+1) - \vec
n(\vec x,t))^2 \label{eq:4.1}\\
&& -
( \vec n(\vec x,t+1) - \vec n(\vec x,t))^2\Bigr]\Bigr\}\nonumber
\end{eqnarray}
When computing a correlation function of $\mu$'s the definition amounts to
replace at the time $t$
the time derivative term of the action
$[\Delta_0\vec n(\vec x,t)]^2$ (the second term in parenthesis of
eq.(\ref{eq:4.1})) by the first term.

We will compute the correlator
\begin{equation}
{\cal D}(x_0) =
\langle \mu_{-q}(\vec 0,x^0) \mu_q(\vec 0,0)\rangle\label{eq:4.2}\end{equation}
i.e. the propagation of a soliton sitting in the origin from time 0 to time
$x_0$.

At large values of $|x_0|$, by cluster property one expects
\begin{equation}
{\cal D}(x_0) \simeq A \exp(-M |x_0|) + \langle \mu_q\rangle^2
\label{eq:4.3}\end{equation}
$\langle \mu_q\rangle\neq0$ in the thermodinamic limit signals spontaneous
breaking of the $U(1)$ symmetry
(\ref{eq:2.19a}).

Our guess is that this spontaneous breaking is related to the phase transition,
and that
$\langle\mu_q\rangle$ is the disorder parameter of the system, dual to the
magnetization in the low temperature phase.

To check that we will compute $\langle\mu_q\rangle$ numerically,
and in particular we will study its behaviour around $\beta_c$. As
a byproduct we shall determine by a finite size scaling analysis
$\beta_c$ and the critical index $\nu$ of the spin correlation
length, and $\beta_c$.

To show that $\mu_q$ actually creates a soliton let us compute ${\cal D}(x_0)$ of
eq.(\ref{eq:4.2}) by use of the definition
(\ref{eq:4.1}) of $\mu_q$
\begin{equation}
{\cal D}(x_0) = \frac{\displaystyle \int d\Omega \exp(-\beta S)
\mu_{-q}(\vec 0,x_0) \mu_q(\vec 0,0)} {\displaystyle \int d\Omega
\exp(-\beta S)} = \frac{Z[S+\Delta
S]}{Z[S]}\label{eq:4.4}\end{equation} According to the definition
(\ref{eq:4.1}) $S+\Delta S$ is obtained from $S$
(eq.(\ref{eq:2.1})) by the replacements at the two time slices
$t=0$ and $t=x_0$
\begin{eqnarray}
t=0 &&
\left(\Delta_0\vec n(\vec x,0)\right)^2 \to
\left( R_q^{-1}\vec n(\vec x,1) - \vec n(\vec x,0)\right)^2\label{eq:sisa}\\
t=x_0 &&
\left(\Delta_0\vec n(\vec x,x_0)\right)^2 \to
\left( R_{-q}^{-1}\vec n(\vec x,x_0+1) - \vec n(\vec x,x_0)\right)^2
\label{eq:sisb}
\end{eqnarray}
$Z[S+\Delta S]$ can be computed by the change of variables $\vec
n(\vec x,1)\to R_q \vec n(\vec x,1)$. As observed at the beginning
of sect.2 this change leaves the measure invariant. The term
$(\Delta_o\vec n)^2$ is restored to the primitive form, but
\[(\Delta_i\vec n)^2\to (\Delta_i R_q \vec n)^2\]
(i.e. a soliton has been created at time $t=1$) and
\[ \Delta_0\vec n(\vec x,2)\to
(R_q^{-1}\vec n(\vec x,2) - \vec n(\vec x,1))\] which translates
the change (\ref{eq:sisa}) at $t=1$.

The construction can be repeated until time $x_0$ is reached, when the
construction produces again a soliton and
\[(\Delta_0\vec n(\vec x,x_0))^2
\to
\left( R_{-q}^{-1}\vec n(\vec x,x_0+1) -
R_{q}\vec n(\vec x,x_0)\right)^2 = (\Delta_0\vec n(\vec x,x_0))^2\]
since $R_{-q}^{-1} = R_q$. This shows that
${\cal D}_q(x_0)$ actually describes a soliton propagating from $t=0$ to
$t=x_0$.
\par\noindent
Instead of $\langle\mu\rangle$, the quantity $\rho =
\frac{d}{d\beta}\ln\langle\mu\rangle$ will be measured\cite{5}.
Since $\langle\mu\rangle_{\beta = 0}$~$=1$
\begin{equation}\mu =
\exp\left[ \int_0^\beta \rho(x) d x\right]\end{equation}
$\rho(\beta)$ is shown in fig.1 for different sizes of the lattice
. At high $\beta$'s the integral defining ${\cal D}(x_0)$ is
gaussian and $\rho$ can be explicitly computed.  The result is
\begin{equation}
\rho = -c_1 L + c_2 \qquad (\beta > \beta_c)\end{equation}
 with $c_1> 0$. The
comparison of this analytic computation to numerical results is
displayed in the figure.
\par\noindent
\begin{minipage}{0.9\textwidth}
\epsfig{figure=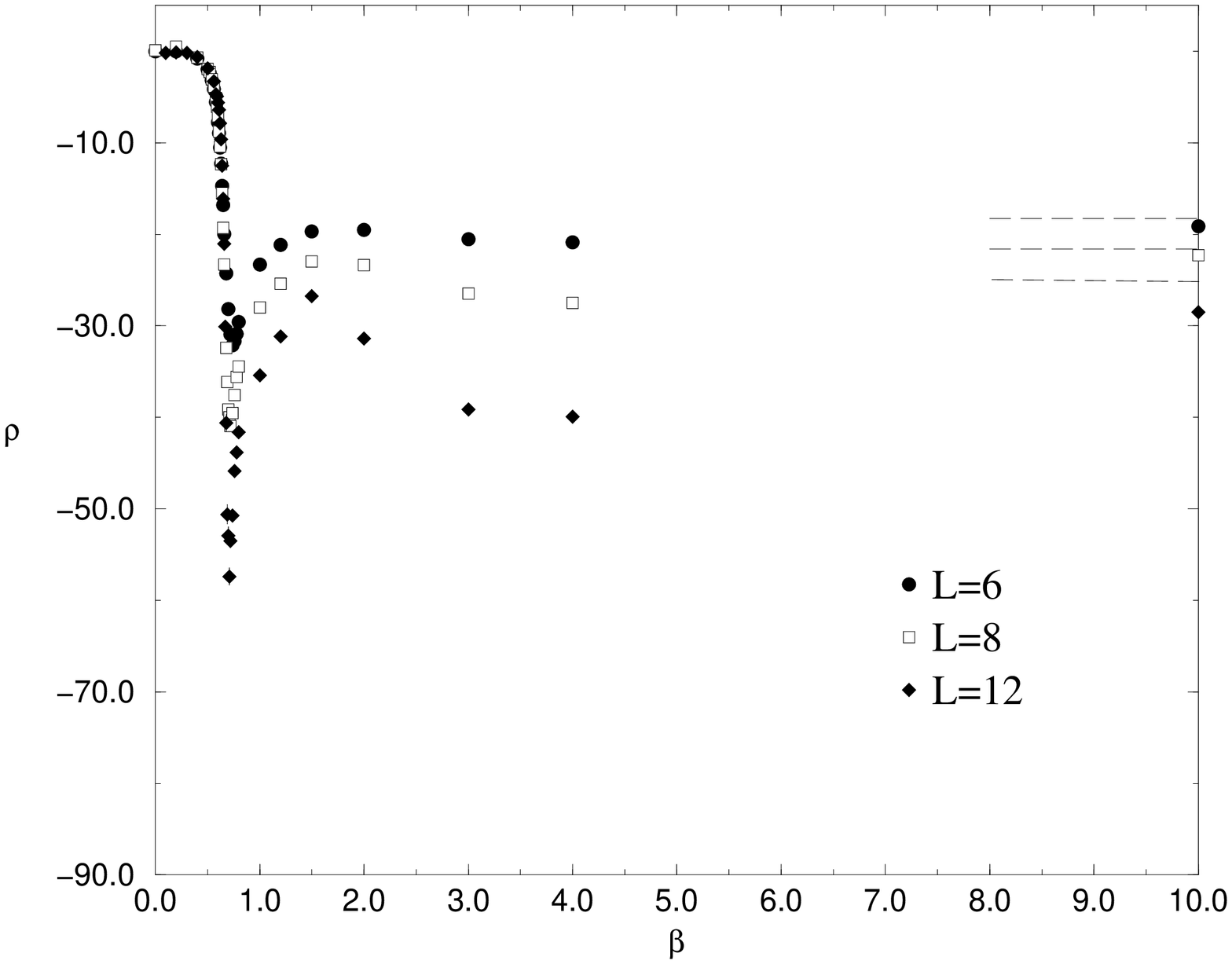,  width = 0.9\textwidth, angle=270}
\vskip0.05in {\bf Fig.1} $\rho$ vs $\beta$. The perturbative
evaluation at high $\beta$ is shown by the dotted lines for
comparison.
\end{minipage}
\vskip0.15in
$\langle\mu\rangle$ is an analytic function of
$\beta$ for any finite $L$, and therefore it cannot vanish in a
region of $\beta$'s without vanishing identically everywhere. Only
in the thermodynamical limit Lee-Yang singularities are produced
and $\langle\mu\rangle$ can be identically zero for $\beta >
\beta_c$, as a honest disorder parameter.

At $\beta<\beta_c$, $\rho$ tends to a finite limit as $V\to\infty$
(fig.2), or $\langle\mu\rangle$ tends to a value different from zero, implying
that $U(1)$ symmetry, eq.(\ref{eq:2.19a}), is spontaneously broken.

Around $\beta_c$ a finite size analysis can be made to determine
$\beta_c$ and the critical index $\nu$. The argument is that, for
dimensional reasons\cite{6}.
\begin{equation} \langle \mu\rangle =
f(\frac{L}{\xi},\frac{a}{\xi})\end{equation} with $\xi$ the
correlation length,$a$ the lattice spacing, $L$ the lattice size.
Near the critical point $\xi$ diverges with a critical index $\nu$
\begin{equation}
\xi \simeq |\beta_c - \beta|^{-\nu}
\end{equation}
\par\noindent
\begin{minipage}{0.9\textwidth}
\epsfig{figure=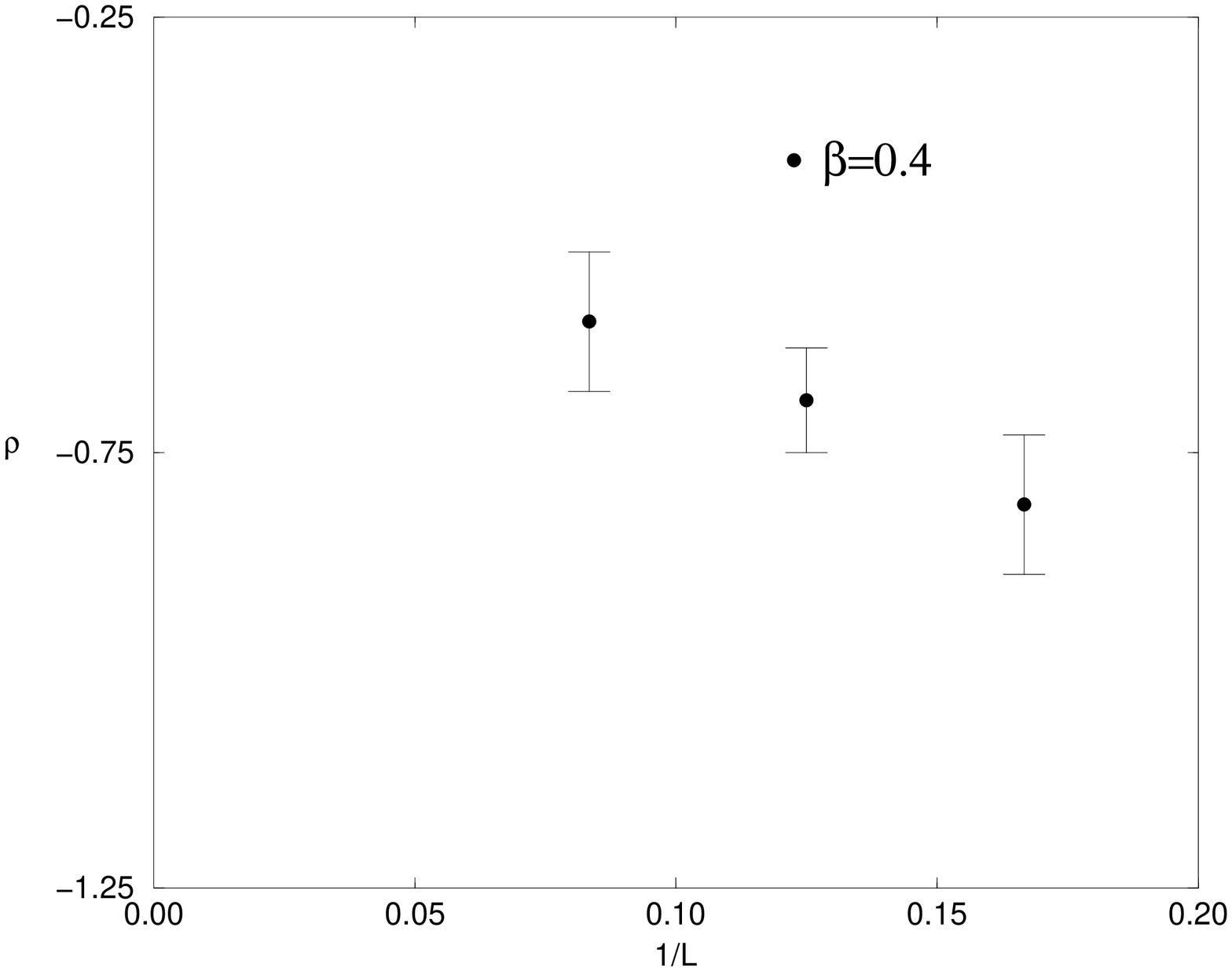,  width = 0.8\textwidth, angle=270}
\vskip0.05in {\bf Fig.2} $\rho$ at high temperature as a function
of $1/L$, ($L$ size of the lattice).
\end{minipage}
\vskip0.15in

 As $\xi$ goes large, the dependence of $\mu$ on
$a/\xi$ can be neglected, and
\begin{equation}
\langle\mu\rangle = f(\frac{L}{\xi},0) =
\Phi(L^{1/\nu}(\beta_c-\beta))
\end{equation}
 This induces on $\rho =
d\ln\langle\mu\rangle/d\beta$ the scaling law
\begin{equation}
\frac{\rho}{L^{1/\nu}} = F(L^{1/\nu}(\beta_c-\beta))
\end{equation}
 Corresponding
to the correct value of $\nu$ and $\beta_c$ determinations of
$\rho L^{-1/\nu}$ coming from lattices of different size should
follow the same curve if plotted versus
$L^{1/\nu}(\beta_c-\beta)$. The quality of scaling is shown in
fig.3. A best fit procedure gives
\begin{equation}
 \nu = 0.70 \pm 0.02\;
[0.704(6)]\qquad \beta_c = 0.695\pm0.003\; [0.6929(1)]
\end{equation}
 For
comparison the official values\cite{14a,17,17a} are shown in
square brackets.
\par\noindent
\begin{minipage}{0.9\textwidth}
\epsfig{figure=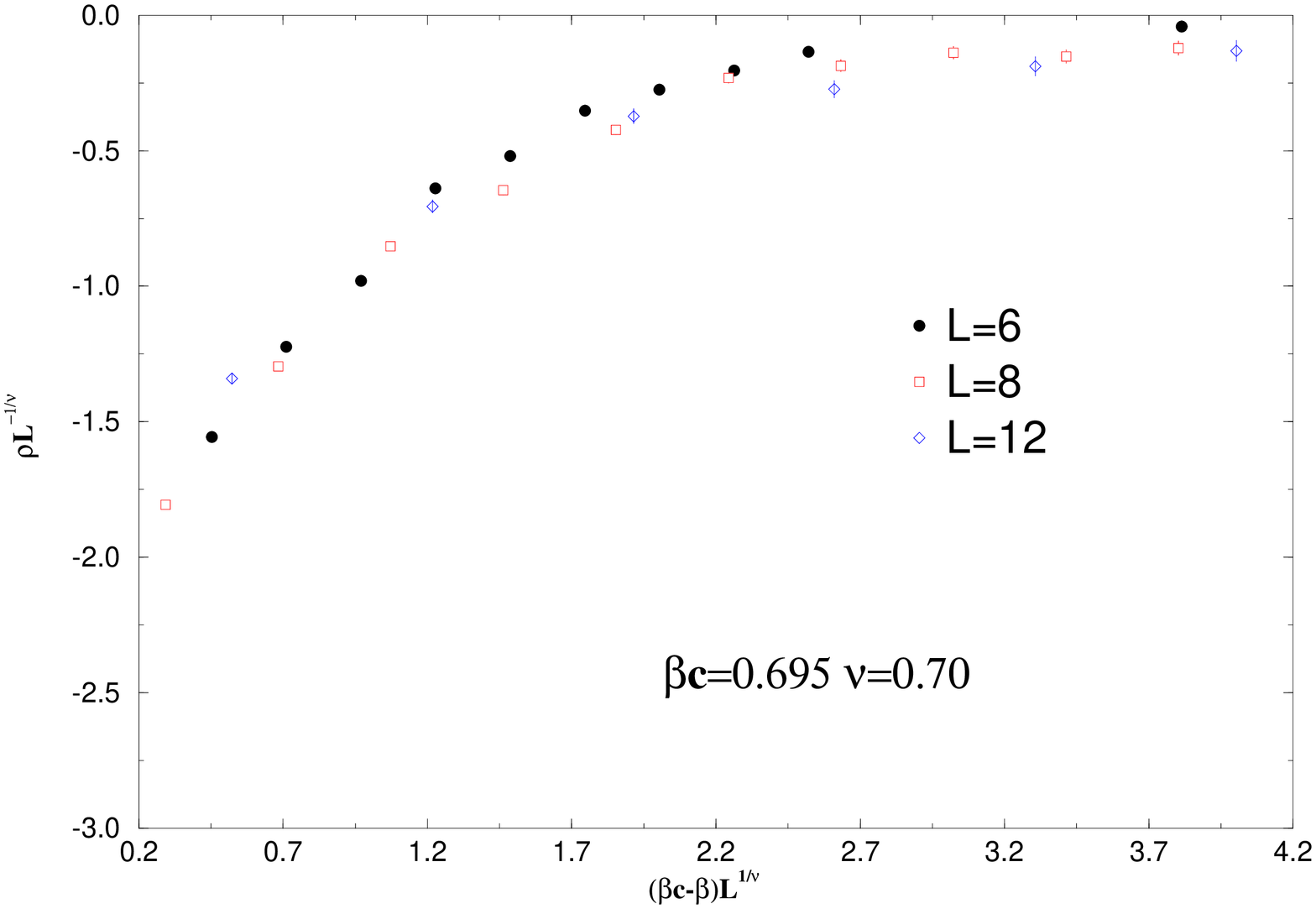,  width = 0.9\textwidth, angle=270}
\vskip0.05in {\bf Fig.3} Finite size scaling at the optimal values
of $\beta_c$ and $\nu$.
\end{minipage}
\vskip0.15in
\section{On the construction of $R_q(x,y)$}
We want now to analyse in detail the definition and the properties of the
rotation $R_q$, introduced in the previous section (eq.(\ref{eq:4.1})).
In principle the model is defined with the boundary condition that
\[\vec
n(x)\to_{|x|\to\infty}\vec n_0\]
to have a finite action.
A time independent rotation $R_q(x,y)$ which creates a vortex when acting on a
uniform  configuration in space
\[ \vec n(\vec x,t) = \vec n_0 = (0,0,1)\]
is known
\begin{equation}
R_q(x,y) = R_z(\theta) R_x(f) R_z^{-1}(\theta) \label{eq:5.1}\end{equation}
with
\begin{equation}
\theta = arctg\frac{(\vec x-\vec y)_2}{(\vec x-\vec y)_1}\qquad
f = f(r),\quad r = \sqrt{(\vec x-\vec y)^2}
\label{eq:5.2}\end{equation}
and with the boundary conditions
\begin{equation}
f(\infty) = 0\qquad f(0) = \pi \label{eq:5.3}\end{equation}
The connection $\vec \omega_\mu$ produced by $R_q$
\[\vec \omega_\mu = \partial_\mu R_q\, R_q^{-1}\]
is, in polar coordinates
\begin{equation}
\vec \omega_0 = 0\qquad
\vec\omega_\theta =
\left(\matrix{\sin\theta\sin f\cr\cos\theta \sin f\cr 1 - \cos
f\cr}\right)\qquad
\vec \omega_r = - f'
\left(\matrix{\cos\theta\cr-\sin\theta\cr 0\cr}\right)
\label{eq:5.4}\end{equation}
The topological charge, as given by eq.(\ref{eq:2.22}), is
\begin{equation}
\frac{1}{4\pi}\oint_{C_1} \vec\omega_\theta\cdot\vec n d\theta +
\frac{1}{4\pi}\oint_{C_2} \vec\omega_\theta\cdot\vec n d\theta =
(1-\cos f(0)) n_3(\vec y) = 1 \label{eq:5.5}\end{equation} where
the line integral is taken on a small circle $C_1$ around the
point $\vec y$ and a circle $C_2$ at $|\vec x | = \infty$. The
first integral gives $(1-\cos f(0)) n_3(\vec y)$, the second
$(1-\cos f(\infty)) n_3(\vec \infty)$. we have chosen $f(0) = \pi$
$f(\infty)=0$. Alternatively one can take $f(\infty) = \pi$ and
$f(0) = 0$, and then the role of the two circles $C_1$ and $C_2$
is interchanged. If we act on a generic configuration $\vec n(\vec
x)$
\[\vec n(x) \to R_q \vec n(x) = \vec n'\]
then we can compute the variation $\delta Q$ of the topological charge produced
by $R_q$
\begin{eqnarray}
\delta Q &=&\frac{1}{4\pi}
\int \vec n'(x)\cdot(\partial_1 \vec n'\wedge \partial_2\vec n')\,d^2 x
- \frac{1}{4\pi}
\int \vec n(x)\cdot(\partial_1 \vec n\wedge \partial_2\vec n)\,d^2 x\nonumber\\
&=&\frac{1}{4\pi}
\int \vec n\cdot(\partial_1 R_q R_q^{-1} \vec n\wedge \partial_2\vec n) d^2 x +
\frac{1}{4\pi}\int \vec n\cdot(\partial_1  \vec n\wedge \partial_2 R_q R_q^{-1}
+\nonumber\\
&+&\frac{1}{4\pi}
\int \vec n\cdot(\partial_1 R_q R_q^{-1} \vec n\wedge \partial_2 R_q R_q^{-1}
\vec n) d^2 x\label{eq:5.6}
\end{eqnarray}
Assuming the form of eq.(\ref{eq:5.1}) for $R_q$ the result is
\begin{equation}
\delta Q = \frac{1}{2}\left[ \vec n(\vec y) \vec\omega_\theta(\vec
y) - \vec n(\infty)\vec\omega_\theta(\infty)\right]
\label{eq:5.8}\end{equation} with $\vec \omega_\theta$ defined in
eq.(\ref{eq:5.4}). If $f(\infty) = \pi$, $f(0) = 0$ $\vec
\omega_\theta(0) = 0$, $\vec \omega_\theta(\infty) = (0,0,2)$ and
$\delta Q = 1$. The rotation $R_q$ changes by 1 the number of
vortices. In our numerical simulations we have first tried a fixed
b.c., $\vec n_b = \vec n_0 = (0,0,1)$. Since, however, the
correlation length tends to be large when approaching the critical
point, the lattice tends to be dominated by surface effects with
very little bulk of the system, for reasonable lattices sizes. An
alternative possibility is to have one single line parallel to the
time axis ($\vec x = \vec y$) on which $\vec n = \vec n_0$ and put
the vortex generated by $R_q$ at the site $\vec y$. This strongly
reduces the boundary effect and makes the lattice much bigger. A
third possibility is to use plain periodic b.c. Then in principle
eq.(\ref{eq:5.8}) as it stands from the continuum, only gives the
correct $\delta Q$ if by chance $\vec n(\vec y) = \vec n_0$. More
than that $R_q$ is defined only if that is true. However if we
change the configuration by putting $\vec n(\vec y) = \vec n_0$
the action only changes by ${\cal O}(1/L^2)$, the sum in
eq.(\ref{eq:5.6}) is not affected by that change to ${\cal
O}(1/L^2)$. Indeed for small values of correlation length the
different boundary conditions described above give the same result
within errors.

We finally compute the connection $\vec \omega_\mu$ and the gauge field
$G_{\mu\nu}(\omega)$ for the configuration of a vortex. We shall refer to a
configuration
\begin{equation}
n_+ = \frac{2 w}{1 + |w|^2|}\qquad n_3 = \frac{1-|w|^2}{1+|w|^2}
\label{eq:5.9}\end{equation}
with $w$ a meromorfic funtion of degree $q$
$w = \prod_1^q \frac{z-a_i}{z-b_i}$.
Let us put
$\vec \omega_\mu = \vec\omega_\mu^{\perp} + f_\mu \vec n$. We get
\begin{eqnarray}
G_{\mu\nu}(\omega) &=& \vec G_{\mu\nu}^{\perp} + \vec G_{\mu\nu}^L
\label{eq:5.10}\\
\vec G_{\mu\nu}^{\perp} &=&
\partial_\mu \vec \omega_\nu^\perp - \partial_\nu\vec \omega_\mu^\perp -
\vec\omega_\mu^\perp\wedge\vec\omega_\nu^\perp\nonumber\\
\vec G_{\mu\nu}^{L} &=& (\partial_\mu f_\nu - \partial_\nu f_\mu)\vec n
\nonumber\end{eqnarray}
Since only $\vec \omega_\mu^\perp$ is defined, we will determine $f_\mu$ in
such a way that $G_{\mu\nu}(\omega) = 0$. The solution is
\begin{equation}
f_1 + i f_2 = i n_3 \frac{1}{\bar w}\frac{\partial \bar w}{\partial \bar z}
\label{eq:5.12}\end{equation}
The topological charge according to eq.(\ref{eq:5.5}) is then
\begin{equation}
Q = \frac{1}{4\pi} \oint_{\Gamma} f_i d x^i = q
\label{eq:5.13}\end{equation}
$\Gamma$ is the contour sum of circles around the singularities of $f$, i.e.
around the poles and zeros of $w$.
\section{The gauge version.}
The gauge version is defined by eq.(\ref{eq:2.4}) in the
continuum.
It is a 2+1 dimensional Georgi-Glashow model
with the size of the Higgs field frozen.
On the lattice it becomes
\begin{equation}
S = \beta \sum_{x,\mu} [D_\mu \vec n(x)]^2 +
\beta' {\rm Tr}\left\{ \Pi_{\mu\nu}\right\}
\label{eq:2.5}\end{equation}
where
\begin{equation}
D_\mu \vec n(x) = \vec n(x+\hat\mu) - U_\mu(x)\vec
n(x)\label{eq:2.6}\end{equation}
and $\Pi_{\mu\nu}$ is the plaquette.
Again a body fixed frame can be defined, and with it $\vec \omega_\mu$, exactly
as in sect.2. Now
\begin{equation}
D_\mu \vec n = \partial_\mu\vec n + ( - g \vec A_\mu + \vec
\omega_\mu)\wedge\vec n \label{eq:3.1}\end{equation}
 while $\vec
\omega_\mu$ and $g\vec A_\mu$ transform both as gauge fields.
Under infinitesimal gauge transformation both $\vec\omega_\mu$
and $\vec A_\mu$ transform as gauge fields
\begin{eqnarray}
g \vec A_\mu &\to & \vec\delta\wedge g\vec A_\mu + \partial_\mu\vec \delta
\label{eq:3.2}\\
 \vec \omega_\mu &\to & \vec\delta\wedge \vec \omega_\mu +
\partial_\mu\vec \delta\nonumber
\end{eqnarray}
The combination $  \vec\omega_\mu - g \vec A_\mu $ is covariant\cite{16}.
In the absence of singularities the field $\vec\omega_\mu$ can be gauged away
by a transformation which brings $\vec n(x)$ to $\vec n_0$, $R^{-1}(x)$, where
$R$ is defined by $\vec n(x) = R(x)\vec n_0$. If the transformation is singular, $\vec
G_{\mu\nu}(x)$ is not covariant but acquires a singular additive term, parallel
to $\vec n$, $\vec G_{\mu\nu}(\omega)$
\begin{equation}
\vec G_{\mu\nu} \to \vec G_{\mu\nu}(\vec A) + \vec G_{\mu\nu}(\vec \omega)
\label{eq:3.3}\end{equation}
The gauge transformation $R^{-1}(x)$ is called an abelian projection. Going to the BFF
makes $\vec\omega_\mu=0$ and $\partial_\mu\vec n = 0$. The $U(1)$ invariance
corresponding to the rotations around $\vec n$ becomes in this frame a $U(1)$
abelian gauge invariance. The corresponding gauge field is
$\partial_\mu A^3_\nu - \partial_\nu A^3_\mu$, or
\begin{equation}
F_{\mu\nu} = \vec n\cdot\vec G_{\mu\nu} -
g(\vec A_\mu \wedge\vec A_\nu)\cdot\vec n \label{eq:3.4}\end{equation}
By use of eq.(\ref{eq:3.4})
\[
g(\vec A_\mu \wedge\vec A_\nu)_3 =
\frac{1}{g} \vec n\cdot(D_\mu\vec n\wedge D_\nu\vec n)\]
and
\begin{equation}
F_{\mu\nu} = \vec n\cdot\vec G_{\mu\nu} -
\frac{1}{g}\vec n\cdot(D_\mu\vec n\wedge D_\nu\vec n)
\label{eq:3.5}\end{equation}
which is the 't~Hooft tensor, and is gauge invariant. Calling
\begin{eqnarray}
j_\mu &=& \partial^\mu F_{\mu\nu}\label{eq:3.6a}\\
j^*_{\mu} &=& \varepsilon_{\mu\alpha\beta} F^{\alpha\beta}
\label{eq:3.6b}\end{eqnarray}
$j^*_{\mu}$ is identically conserved, since in the abelian projected frame
$F_{\mu\nu}$ is a curl.
In the limit $g\to 0$, $g j^*_\mu$ coincides with the current (\ref{eq:2.19}).
The above analysis shows that the Heisenberg model can be seen as
the $g\to 0$ limit of a Higgs model. The abelian projection is the
transformation to BFF. Its singularities depend on the Higgs field
configurations: the gauge field for the connection $\vec \omega_\mu$
is present both in the gauged and in the simple version of
the model.
\section{Concluding remarks}
A disorder parameter $\langle  \mu\rangle$ has been defined for
the phase transition of demagnetization in the 3d Heisenberg
model. $\langle  \mu\rangle$ vanishes in the magnetized phase, and
is non zero in the disordered phase, signalling condensation of
vortices in the vacuum. The vortices are the instantons of the 2d
version of the model. The critical index $\nu$ and the transition
temperature $\beta_c$ can be determined by a finite size scaling
analysis, and they  agree with the values obtained from the side
of ordered phase. Duality implies a non trivial topological
structure of the model. Vortices can be viewed as gauge
singularties resulting from the abelian projection. In fact they
have a physical role on the dynamics of the system.

\end{document}